# Multi-step nucleation of nanocrystals in aqueous solution

N. Duane Loh[1, 2], Soumyo Sen[3], Michel Bosman[4, 5], Shu Fen Tan[6], Jun Zhong[1, 2]
Christian A. Nijhuis[6, 7]†, Petr Král[3, 8]†, Paul Matsudaira[2, 9], Utkur Mirsaidov[1, 2, 7, 10]†

1. Department of Physics, National University of Singapore, 2 Science Drive 3, Singapore, 117551.
2. Center for Bioimaging Sciences, Department of Biological Sciences, National University of Singapore, 14 Science Drive 4, Singapore, 117543.
3. Department of Chemistry, University of Illinois at Chicago, Chicago, Illinois 60607, USA.
4. Institute of Materials Research and Engineering, A*STAR (Agency for Science, Technology and Research), 2 Fusionopolis Way, Singapore 138634.
5. Department of Materials Science and Engineering, National University of Singapore, 9 Engineering Drive 1, Singapore 117575.
6. Department of Chemistry, National University of Singapore, 3 Science Drive 3, Singapore, 117543.
7. Centre for Advanced 2D Materials and Graphene Research Centre, National University of Singapore, 6 Science Drive 2, Singapore 117546.
8. Department of Physics, University of Illinois at Chicago, Chicago, Illinois 60607, USA.
9. MechanoBiology Institute, National University of Singapore, 5A Engineering Drive 1, Singapore 117576.
10. Nanocore, National University of Singapore, 4 Engineering Drive 3, Singapore 117576.

**Abstract.**
**Nucleation and growth of solids from solutions impacts many natural processes and are fundamental to applications in materials engineering and medicine. For a crystalline solid, the nucleus is a nanoscale cluster of ordered atoms, which forms through mechanisms that are still poorly understood (*1*, *2*). These mechanisms have important consequences on the morphology and nucleation rates of the resultant crystals (*3*, *4*) but it is unclear whether a nucleus forms spontaneously from solution in a single step or through multiple steps. Using *in-situ* electron microscopy, we observe and quantify how gold and silver nanocrystals nucleate from a supersaturated aqueous gold and silver solution in three distinct steps: (I) spinodal decomposition into solute-rich and solute-poor liquid phases, (II) nucleation of amorphous gold nanoclusters within the gold-rich liquid phase, followed by (III) crystallization of these amorphous clusters. Our *ab-initio* calculations on gold nucleation suggest that these steps might be associated with strong gold-gold atom coupling and water-mediated metastable gold complexes. The understanding of intermediate steps in nuclei formation has important implications for the formation and growth of both crystalline and amorphous materials.**

Nucleation is commonly described by the Classical Nucleation Theory (CNT), in which a nascent phase, termed *nucleus,* emerges from solution in a single-step (*1*, *5–7*). Although CNT successfully describes various phenomena, such as the condensation of water droplets from vapor (*8*), it fails for more complex systems. For example, CNT predicts homogeneous nucleation rates for lysozyme (*3*) and ice (*9*) that are at least ten orders of magnitude slower than experimentally measured rates. To account for the large discrepancies in crystallization rates (*1*, *3*, *10*), most non-classical nucleation models introduce an additional step preceding nucleation (*5*, *10*). One such mechanism involves spinodal decomposition (*11*) of two-component solutions to form solute-rich and solute-poor liquid phases followed by the formation of nuclei in the solute-rich liquid phase. This multi-step mechanism has been proposed for the nucleation of $CaCO_3$ in water (*12*, *13*), lysosome crystals in water (*5*), and others (*14*). To develop models for nucleation that deviate from CNT, direct experimental probing of the nanoscale dynamics of alternative nucleation pathways is necessary.

Whilst others have studied post-nucleation crystal growth at the nanometer-scale (*15–17*) using recent advances in *in situ* Transmission Electron Microscopy (TEM) (*15, 18*), here we extend this TEM method to directly observe each step in the nucleation pathway of gold nanoparticles in water. We choose to investigate the nucleation pathway of gold and silver from water because previous studies of gold crystallization predicted that a dense liquid phase should nucleate from a supersaturated solution (*19, 20*) (Section SI11A), and that nanocrystals form rapidly from this dense phase (*21, 22*) (Section SI11B), likely via spinodal decomposition (*20*) (Section SI12). These studies, however, did not uncover the dynamics of the phases preceding nucleation. To test these predictions, we used *in situ* TEM to follow the nucleation dynamics of gold from an aqueous 1 mM $HAuCl_4$ solution, sealed in a liquid cell (*15, 23, 24*) comprising two ultrathin (~14 nm) $SiN_x$ membranes separated by 200 nm spacers (Section SI1) at a frame rate of 10 Hz. When irradiated by electrons, the bulk of aqueous solution recedes, leaving a ~30 nm-thick aqueous film (*25*) (details in Sections SI2-5) in which the gold-ions are reduced *in situ* by the electron beam and eventually form gold nanoparticles. This approach enables us to track the nucleation of these gold nanoparticles with insignificant thermal perturbation (Section SI2) and by eliminating chemical reducing agents.

Real-time imaging (Figure 1A) shows two intermediate states of gold nucleation (Supporting Video 1). Between the initial homogeneous solution of gold (left panel) and the final crystalline solid nanoparticle (right panel) two sequential states (middle panels) appear: a two-phase aqueous mixture and a non-crystalline cluster. The initial phase separation can be explained by the well-documented fact that high-energy electrons (*26*) generate solvated electrons in the aqueous gold solution leading to rapid reduction of $Au^{3+}$ ions to $Au^0$ (*26*) (Sections SI2 and SI4). Other short-lived species created during radiolysis of water do not promote gold nucleation owing to their lower redox potentials (Section SI4). Qualitative molecular dynamics simulations (Figure 1B; Section SI14) confirm that gold nanocrystals nucleate from the supersaturated aqueous solution of $Au^0$ in three steps: (I) the solution demixes into gold-poor and gold-rich aqueous phases by spinodal decomposition, (II) amorphous gold nanoclusters arise from the gold-rich phase, and (III) the amorphous nanoclusters crystallize. These three steps are not specific for gold nucleation but also appear when forming silver nanocrystals in water (Figure 1C and Section SI15), implying that multi-step nucleation likely applies to other noble-metal nanoclusters, an important class of catalysts. We present our quantitative analysis methods below, focusing on gold as an example.

To identify the mechanisms of gold nanoparticle formation, we examined the initial steps of the formation of the gold-rich phase and its transition into nanoclusters of diameters between 10 and 38 Å (Figure 2A). From mass measurements by STEM microscopy of amorphous nanoclusters (Figure S4A), we estimate that the gold concentration in the pre-nucleation gold-rich aqueous phase is ~4 M. Calculating backwards by redistributing the atoms in tracked amorphous nanoclusters (assumed to have the density of bulk gold) uniformly over a 30 nm-thick liquid film (Section SI5), we estimate that the initial homogeneous gold concentration must have been 0.2-0.6 M (Figure 2B) or 200-600 times more concentrated than the initial 1 mM solution (Section SI5, and motivated by Section SI7). This increase in $AuCl_4^-$ concentration is consistent with the distribution of counter-ions near the positively-charged $SiN_x$ surface (Section SI7). Similar to radiolysis-induced silver nucleation in water (*16*), our system achieves its threshold induction dose for nucleation sooner with higher electron fluxes, thereby also accelerating the formation of nanoclusters (Figure 2B). These nucleation rates are consistent with those from previous observations of gold nucleation (Section SI11A).

These amorphous nanoclusters display a range of dynamics (Section SI8) including diffusion, rotation, coalescence, and shrinking/growing *via* Ostwald ripening (*27*) or accretion (*28*). By tracking the clusters, we calculate an effective planar diffusion coefficient $D_{MSD}$ of about 5-10 $Å^2 s^{-1}$ (Figure 2C) (*29*), which is nine orders of magnitude smaller than nanoparticle diffusion coefficient in bulk solution (Sections SI7 and SI10). The suppressed diffusion is consistent with previous studies, which is commonly attributed to the interaction of nanoparticles with the membrane surface(*15*, *29*, *30*), and is the reason why we are able to follow all steps of the nucleation process at our experimental time scales of 10 frames per second. From $D_{MSD}$ we estimate an average translational blur of ~ 1.4 Å between images of a 10 Å-radius gold nanocrystal separated by 100 ms, which still allows us to resolve its 2.35 Å lattice spacing (examples in Figure 3).

Since our TEM and cameras have the spatial and temporal resolution more than sufficient to determine the gold nanoclusters' crystallinities, we measured a crystallinity score (Figure 3) from a Fourier parameter around spatial frequencies $k = 2\pi/2.35$ Å$^{-1}$ (elaborated in Section SI9) of 46×46 Å$^2$ regions centered around each nanocluster. This score is sensitive to diffraction contrast from a gold nanocrystal's Face-Centered-Cubic (FCC) {111} lattice spacing. Of the 74 gold nanoclusters that were tracked for Figure 3, 20 showed statistically significant crystallinity above background contrast (with 99.9% confidence) for at least 100 ms during a 5 s observation window as detailed in Section SI9. The remaining 53 nanoclusters are either amorphous, have crystallinities below our detection threshold, or never rotate their lattice planes into view for detection. The last possibility is less likely since these nanoclusters move (Figure 2C) and rotate substantially during the 5 s observation window (*29*). Notably, a meagre 0.8% of the 12,229 views of these 74 nanoclusters show significant crystallinity (observations above the line in Figure 3). This crystalline fraction only increases to 4% for the 620 views of nanoclusters when their radii are greater than 15 Å. In later frames the non-crystalline nanoclusters become crystalline (Figure 1A), suggesting that an amorphous nanocluster directly precedes a crystalline nanoparticle.

The theory of spinodal decomposition explains how small density fluctuations can cause a homogeneous solution to demix into metastable high-entropy gold-poor and low-enthalpy gold-rich aqueous phases. Two features in Figures 1A and 2A are characteristic of spinodal decomposition. First, the initial homogeneous concentration of gold in Figure 2B is far higher than the room-temperature saturation concentration of gold in water (~10$^{-12}$ M binodal, from (*31*)). Hence, it is plausible that our gold-water binary system is deep within its spinodal region. Second, diffusion-limited spinodal structures of feature size $L$ typically appear on time scales $\tau_{sp} \sim L^2/D_{Au}$ (*32*), where $D_{Au}$ is the molecular diffusion of gold atoms (~30 ± 6 Å$^2$s$^{-1}$, extrapolated from Figure 2C using the Stokes Einstein diffusion equation Eq. S25). This is consistent with the local gold-rich phases that appear within $\tau_{sp} \sim 10$ s of observation, and indeed have features measuring 1.6–2.0 nanometers as shown Figure 2.

To explain how the gold-rich aqueous phase forms and breaks down in Figures 1A and 2A, we performed *ab-initio* calculations of a hydrated gold atom pair (Figure 4). This hydrated atom pair becomes ionized when brought closer together: the left gold atom plus two nearby water molecules form a linear cationic coordination complex, [Au(H$_2$O)$_2$]$^{+1}$, while the right gold atom becomes an anion surrounded by a simple hydration shell. Other (square planar and linear) complexes involving chloride and hydroxide ligands may also participate, depending on pH (*26*, *33*) (Section SI2). For nanoclusters to form inside the gold-rich aqueous phase, pairs of gold atoms within it must be partially dehydrated. In our calculations (Figure 4), this dehydration is delayed by a 7.6 kcal/mol (12.9 $k_BT$) energy barrier required to breakdown the linear cationic complex (close to the gold anion). Such a barrier metastabilizes a fluidic network of hydrated ionized complexes in the gold-rich phases (Sections SI13 and SI14) that gradually shed their waters to form amorphous nanoclusters, as illustrated in our qualitative molecular dynamics simulations (Figure S25). Most of the amorphous nanoclusters do not redissolve back into the gold-rich aqueous phases as shown in Figure SI10, which agrees with the high-energy barrier for dissolution in our calculations (31.8 kcal/mol or 53.9 $k_BT$) in Figure 4.

Our experiment allows us to follow the evolution of intermediates present in the current nucleation. Typically, during spinodal decomposition, the surface tension between the demixed phases causes spinodal structures to coarsen and grow (*34*) (also see Figure SI21). These spinodal gold structures shown in Figures 1A are unstable and condense into amorphous nanoclusters. This shrinking suggests that a second (condensed) phase transition follows spinodal decomposition to produce amorphous nanoclusters. The amorphous organization of the ~2-3 nm particles is consistent with steeply decreasing melting points of gold nanoparticles when their radii shrink below 5 nm (*21*, *35*). Below this size, gold clusters containing twenty or fewer atoms adopt a variety of non-FCC structures (*36*). In comparison, Li *et al*. show room temperature platinum nanoparticles, which like gold is also a noble metal, become increasingly disordered when their diameters fall below 2 nm (*37*). These studies are compatible with how most gold clusters, whose diameters are smaller than 3 nm, that appear during multi-step nucleation for Figure 3A are structurally amorphous. Our experimental findings show how gold-rich spinodal structures condense into amorphous nanoclusters, which then crystalize into nuclei that are large enough to be stable and support nanoparticle growth.

The spinodal gold structure is an initial step in nucleation and requires the electron beam to initiate gold cluster formation by reducing gold ions in the solution. However, we do not believe that the electron beam used in our experiments has affected the true character of multi-step nucleation for the following four reasons. First, the energy transferred by electrons heats the sample by less than 5 K (Section SI3), which rules out beam-induced melting of intermediate or final state products. Second, evidence of spinodal decomposition was previously reported for gold nucleation even in the absence of ionizing radiation in considerably thicker liquid layers (*20*). The fact that we resolved this transient spinodal decomposition suggests that our imaging conditions did not change the nature of the intermediate steps during nucleation. Third, previous studies of radiolysis-induced nucleation of metallic salts (*26*, *38*) did not conclude that increasing radiation doses would fundamentally alter the underlying chemistry driving nucleation (Section SI4). This is consistent with our observations, where spinodal decomposition always precedes nanocluster formation even when we quadrupled the average electron flux (Figure 2) or increased the peak flux by seven orders of magnitude using scanning electron probe (Section SI6). Finally, the nucleation rates reported here are consistent with experiments without ionizing radiation or predicted from numerical simulations (Section SI11 and SI12), indicating that the mechanisms underlying nucleation were largely unaffected by the electron beam. Overall, we found no compelling evidence that the electron beam, aside from initiating nucleation, has fundamentally affected the observed mechanism of multi-step nucleation.

Our observations clearly show that gold nanocrystals can nucleate from a super-saturated aqueous solution *via* a three-step mechanism: spinodal decomposition, solidification, and crystallization. The amorphous nanocluster is a direct precursor for the crystalline nucleus in solution and we anticipate that this sequence could be common in crystallization. If so, then the amorphous nanocluster would be an intermediate common to formation of both crystalline and amorphous solids from solution. More generally, this ability to observe pathways in crystal nucleation will impact research on catalysis, nanoparticle synthesis and structural biology. Our findings show that multi-step nucleation pathways are feasible, quantifiable, and will help to develop better and broadly applicable nucleation models.

**Acknowledgement**: This work was supported by the Singapore National Research Foundation's Competitive research program funding (NRF-CRP9-2011-04). CAN and MB would like to acknowledge support from No. NRF-CRP 8-2011-07, UM and PM acknowledge support from MBI, and CBIS. NDL thanks the support of the Lee Kuan Yew Endowment Fund. The work of P.K. was supported by the NSF DMR grant No. 1309765 and by the ACS PRF grant No. 53062-ND6.

**Figures**

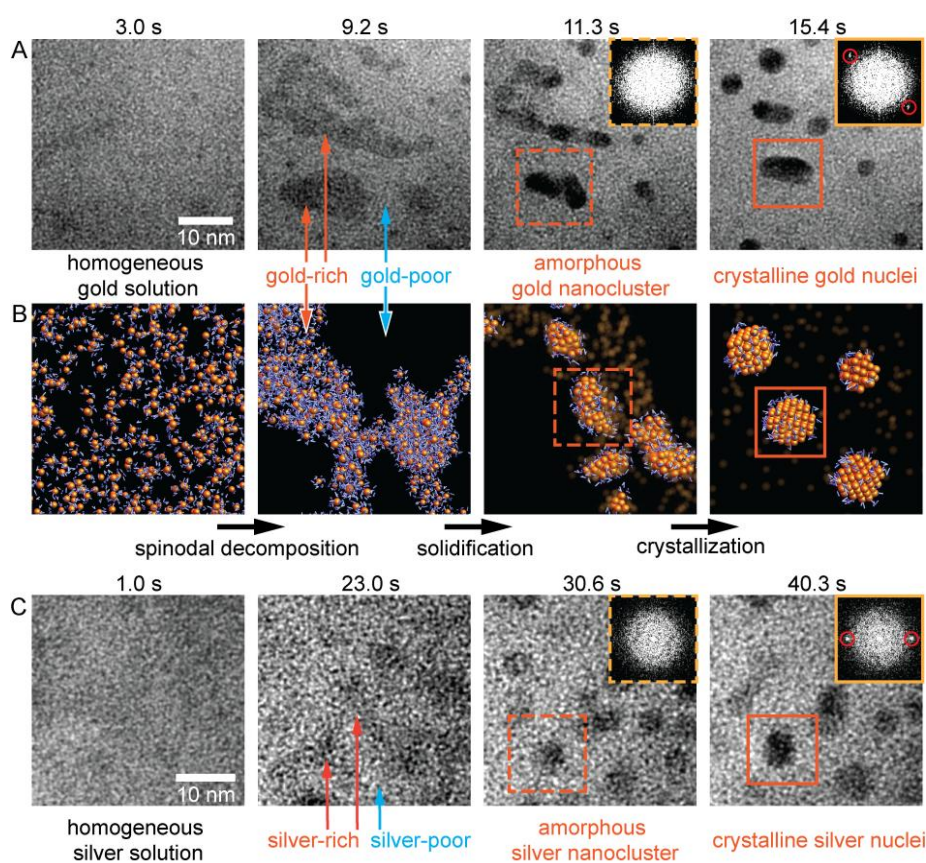

**Figure 1**. **Proposed three-step pathway for gold (Au) and silver (Ag) nucleation.** (**A**) A series of TEM images showing the intermediate steps in nucleating gold nanocrystals from a supersaturated aqueous $Au^0$ (Supporting Video 1). From 3.0-9.6 s, supersaturated $Au^0$ solution spontaneously demixes into Au-poor and Au-rich liquid phases (lighter and darker regions, respectively) *via* spinodal decomposition. Amorphous Au nanoclusters emerge from Au-rich phases that then crystallize. Insets show Fourier transforms of cropped square regions (orange) with the Au{111} FCC reciprocal lattice spacing circled in red. (**B**) Schematic of the proposed steps in nucleation (gold as orange spheres, with surrounding water as blue bent lines) discussed in the text. (**C**) A TEM image sequence where silver nanocrystals nucleate from aqueous $Ag^0$ with the same intermediate steps as gold in (A). Here, a supersaturated $Ag^0$ aqueous solution demixes into Ag-poor and Ag-rich liquid phases (1.0-23.0 s), where the latter develops amorphous Ag nanoclusters (30.6 s) that crystallize. Insets show Fourier transforms of cropped regions (orange) with visible Ag{111} lattice spacing circled in red.

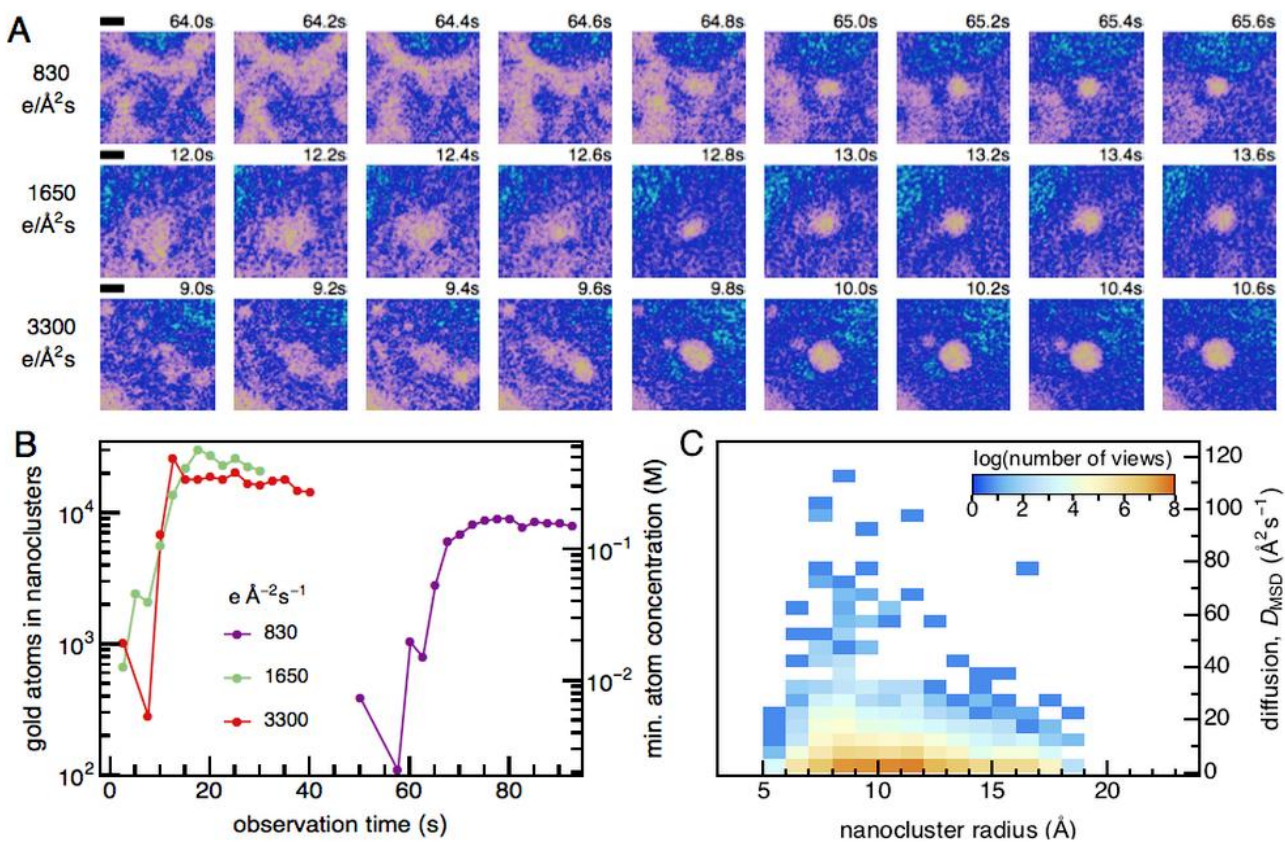

**Figure 2. Amorphous gold nanoclusters appear from gold-rich spinodal structures.** (**A**) High resolution detail of three gold nanoclusters forming from spinodal structures at three TEM electron fluxes, labeled with the duration of TEM electron-irradiation and a 2 nm scale bar. The boundaries between gold-poor (dark) and gold-rich (light) regions are enhanced in this false color image (details in Section SI8). Compact boundaries of nascent gold nanoclusters appear after 65, 12.8 and 9.8 s of observation at electron fluxes 830, 1650 and 3300 e $Å^{-2}$ $s^{-1}$ respectively (Supporting Video 2, 3 and 1 respectively). (**B**) Estimated number of gold atoms in nanoclusters assumed of bulk gold density, when imaged with three electron fluxes. Since nanoclusters diffuse slowly, this number gives a lower bound to the initial homogeneous concentration of atomic gold dissolved in a 30 nm-thick liquid film (thickness estimated in Section SI5). (**C**) Histogram of nanocluster radius versus planar diffusion coefficients. This was computed from more than 15,000 consecutive pairs of TEM frames at the three electron fluxes in (B), for 74 nanoclusters that we tracked for at least 5 s each. Comparatively, the diffusion coefficient of a 1 nm-radius spherical nanocluster in room temperature bulk water is ~ $10^{10}$ $Å^2$ $s^{-1}$ (Section SI10).

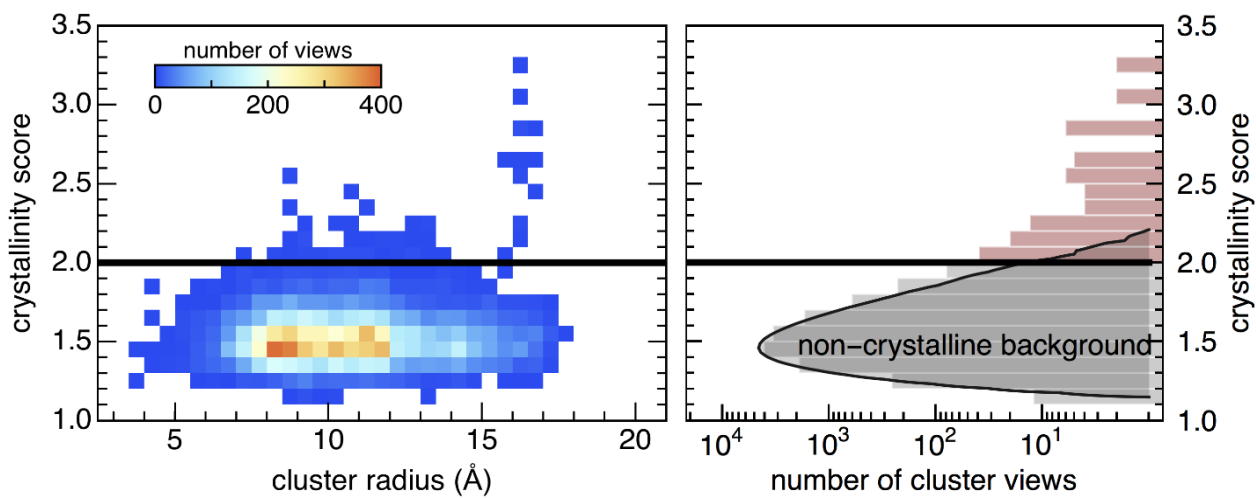

**Figure 3. Crystallinities of 74 gold nanoclusters. Left:** crystallinity score versus cluster radii of 74 gold nanoclusters; at least 50 views on each cluster, totaling to 12,229 views. **Right:** histogram of these crystallinities, superimposed with the histogram of baseline scores for 200,000 background images comprising only silicon nitride and water (black curve, see Section SI9). Of the 74 nanoclusters, 20 (27%) show statistically significant crystallinities above 2.0 (red bars) for at least 100 ms during observation windows longer than 5 s.

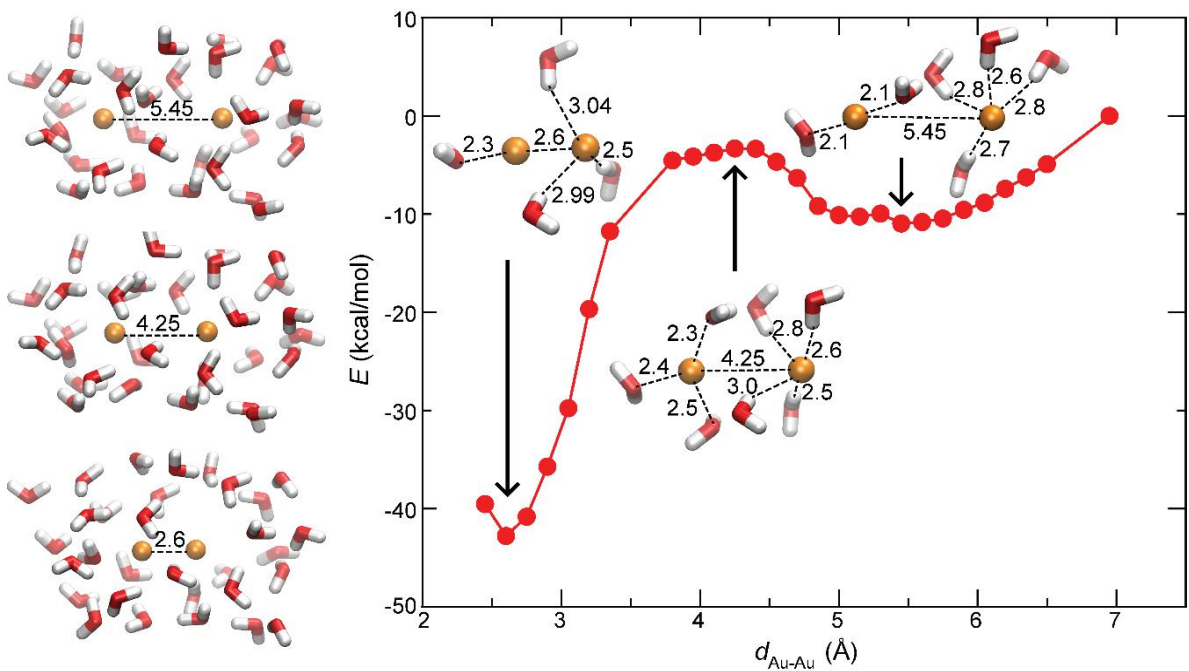

**Figure 4. Ab-initio calculations of a hydrated pair of gold atoms. Left:** two gold atoms (gold spheres) were held at separation $d_{Au-Au}$ while the configurations of thirty-one surrounding water molecules (oxygen and hydrogen as red and white segments respectively) were energetically optimized. We changed the distance between two gold atoms. Out of thirty-one water molecules, nineteen are totally free whereas the oxygen atoms of remaining twelve water molecules were fixed. **Right:** the computed ground state energy of the hydrated gold system shows two energy minima when $d_{Au-Au}$ is 5.45 Å and 2.60 Å, whose energies differ by 31.8 kcal/mol. The metastable state at $d_{Au-Au}$ = 5.45 Å has to overcome an energy barrier of 7.6 kcal/mol by expelling intervening water molecules to access a denser Au-Au packing when $d_{Au-Au}$ = 2.60 Å.